\documentclass[letterpaper]{article} \usepackage[preprint]{aaai2027}  \usepackage[hyphens]{url}  \usepackage{graphicx}    \usepackage{natbib}  \usepackage{caption}   \usepackage{algorithm}
\usepackage{algorithmic}
\usepackage{amsmath,amssymb,amsfonts}
\usepackage{bm}
\usepackage{amsthm}
\usepackage{multirow}
\usepackage{pifont}
\newcommand{\cmark}{\ding{51}}
\newcommand{\xmark}{\ding{55}}
\usepackage{array}
\newcolumntype{C}[1]{>{\centering\arraybackslash}p{#1}}
\usepackage{newfloat}
\usepackage{listings}
\DeclareCaptionStyle{ruled}{labelfont=normalfont,labelsep=colon,strut=off} 
\floatstyle{ruled}
\newfloat{listing}{tb}{lst}{}
\floatname{listing}{Listing}

\usepackage{booktabs}

\title{High Frequency Safeguard and Weighted Objective Scoring Bandwidth Extension}
\author{Tarikul Islam Tamiti, Sajid Fardin Dipto, Anomadarshi Barua}
\affiliations{}

\begin{document}

\maketitle

\begin{abstract}

Speech Bandwidth Extension (BWE) reconstructs missing high frequencies from a band-limited signal. Please note that perceptually important information is often concentrated in short high-frequency (HF) events. Therefore, an effective BWE requires both accurate HF generation and explicit control over where, how often, and in what spectral form HF energy should be produced. To address the above HF supervision, we present \textbf{HFSWO}, a compact complex-spectral Swin-based dual-stream adversarial framework with two newly proposed specialized discriminators: \textbf{1.} \textbf{High-Frequency Safeguard Discriminator (HFSD)} converts excessive framewise HF-energy ratios into a differentiable safeguard-constraint penalty optimized through a projected primal--dual augmented-Lagrangian update; \textbf{2.} \textbf{Weighted Objective Scoring Discriminator (WOSD)} learns an interpretable convex objective scoring over spectral flatness, centroid, and rolloff. Our rigorous experiments on VCTK English and MLS French datasets across multiple frequency ranges, including noisy conditions, show significantly improved perceptual scores with  74.6\% fewer parameters than the existing baseline AP-BWE, establishing a new baseline in the BWE task in multilingual and multispeaker setups.

\end{abstract}

\section{\textbf{Introduction and Related Work}}
\label{sec:intro}

Speech Bandwidth Extension (BWE) reconstructs wideband speech signals from observations whose upper frequencies are removed by acquisition, transmission, storage, or codec constraints. Therefore, BWE is relevant to telephony, archival restoration, automatic speech recognition, and speech enhancement pipelines. Unlike conventional denoising,  BWE must infer spectral content that is not directly observed. {Therefore, the mapping from a narrowband input to its missing high-frequency (HF) region is a one-to-many mapping problem, and an improper BWE model can produce a spectrally plausible output that might be perceptually incorrect.}

\textbf{Research gap:} Please note that there are clear distinctions between HF events for vowels and fricatives because voiced sounds are predominantly harmonic and shaped by vocal-tract resonances, whereas unvoiced consonants/fricatives contain sudden bursts, and rapid spectral movement \cite{titze2008nonlinear}. Therefore, steady voiced frames generally exhibit limited HF-energy ratios, but fricatives show localized and phoneme-dependent strong HF energy. Moreover, the majority of HF events (i.e., chirps, spikes, or boundary discontinuities)  occupy only a small fraction of an utterance. Therefore, supervision dominated by global spectral averages in recent work {\color{black}\cite{liu2022neural, lee2021nu, lee2025wave, wang2021towards, lu2024towards, kim2023hifi++, hauret2023eben, mandel2023aero,abreu2024aeromamba}} may favor overly smooth outputs. Conversely, blind HF amplification can introduce hiss, over-bright voiced segments, and spurious consonantal cues. \textit{Therefore, effective BWE  requires both accurate HF generation and explicit control over where, how often, and in what spectral form HF energy should be produced.} To the best of our knowledge, no work in the literature provides dedicated HF supervision to suppress sustained over-boost without erasing short consonant bursts by evaluating the structure of generated HF content rather than its total energy alone.

\begin{figure*}[t!]
  \centering
\includegraphics[width=0.98\textwidth,height=0.24\textheight,keepaspectratio]{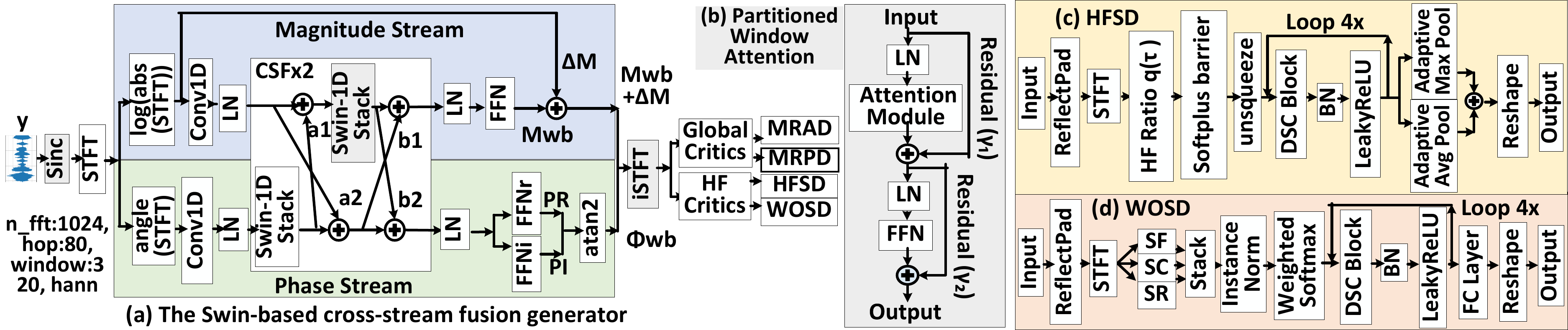}
  \caption{{\color{black} Architecture of (a,b) Swin-1D stacks, CSF, and newly proposed discriminators - (c) HFSD and (d) WOSD.}}
  \label{fig:overall_architecture}
  \vspace{-1em}
\end{figure*}

\textbf{Contribution:} To close the above research gaps, we design \textbf{HFSWO} (\textbf{H}igh-\textbf{F}requency \textbf{S}afeguard and \textbf{W}eighted-\textbf{O}bjective) BWE framework, where two specialized discriminators are introduced, for the first time, to explicitly control over where, how often, and in what
spectral form HF energy should be produced to suppress unintended HF overboosting. 

The first specialized discriminator, \textbf{High-Frequency Safeguard Discriminator (HFSD)}, focuses on excessive framewise HF-energy ratios. It uses a smooth barrier and a projected primal--dual augmented-Lagrangian loss to implement a soft safeguard constraint, where persistent over-boost is penalized, but brief high-energy events can remain possible. This formulation is inspired by the chance-constrained \cite{erdougan2006ambiguous} role of safeguard constraints in controlling the frequency of undesirable outcomes rather than forcing every frame to satisfy a rigid bound.

The second specialized discriminator, \textbf{Weighted Objective Scoring Discriminator (WOSD)}, evaluates the spectral form of reconstructed HF content rather than its energy alone. It computes spectral flatness, normalized centroid, and 85\%-rolloff and learns a convex combination of these criteria. This provides differentiable supervision over noisiness, brightness, and spectral extent. This design is inspired by the multi-criteria principle \cite{keeney1993decisions}, where distinct quality attributes should be represented explicitly rather than collapsed into a single hand-selected statistic. Together, HFSD and WOSD encode complementary phoneme-informed objectives: HFSD discourages sustained upper-band over-boost, while WOSD evaluates where and how HF content is distributed. 
Our  HFSWO achieves the best perceptual performance (NISQA-MOS) while remaining better across other objective metrics (LSD, STOI, PESQ, SI-SNR, WER; see Section \ref{subsec:EvaluationMetrics} for full forms) on English VCTK and MLS French datasets in multilingual and multispeaker setups. We further show the strong robustness of HFSWO in noisy conditions, and the subjective evaluation favors HFSWO over other baselines. 

Most importantly, we show that these gains are obtained with substantially 74.6\%  fewer parameters and 63.6\% less MACs and FLOPs compared to the most competitive baseline AP-BWE \cite{lu2024towards}. We show that this efficiency comes from our compact one-dimensional (1D) Swin-based generator and our newly designed discriminators. \textit{Therefore, for the first time, this paper demonstrate that dedicated HF supervision can be used as effective inductive biases for designing parameter-efficient BWE frameworks.}

\section{Architecture of HFSWO}
\label{sec:overview}

As adversarial training provides better reconstruction capability \cite{lu2024towards, kumar2019melgan, kong2020hifi, kim2023hifi++,tamiti2025nldsi, hauret2023eben}, we design HFSWO as adversarial networks. The overall architecture is shown in Fig. \ref{fig:overall_architecture}.

\subsection{Design of the Generator} \label{subsec:gen}

As accurate reconstruction depends on both phase and magnitude \cite{yin2020phasen}, we design HFSWO's generator to process magnitude and phase in two parallel streams after the Short-Term Fourier Transformation (STFT) of the narrowband input signal ${y}$. Both streams pass through 1-D convolution (Conv1D) and Layer Normalization (LN). 
\textbf{i) Swin 1-D Stack:} In the generator, the central module is a pair of Swin-1D stacks, which use 1-D Swin-based \cite{liu2021swin} shifted windows for capturing both local and global acoustic cues efficiently. Unlike standard global attention by transformer \cite{ashish2017attention}, our Swin-1D stacks perform attention within windows of length $W$ and alternate the offsets $\{0,\lfloor W/2\rfloor\}$ across successive layers. The alternation between regular and shifted windows expands the effective receptive field across window boundaries while avoiding the quadratic complexity of global attention. Thus, a sequence of length $T$ requires $O(TW)$ attention complexity rather than $O(T^2)$, providing better efficiency for our task.

\textbf{ii) Cross Stream Fusion (CSF):} The CSF complements Swin's efficient attention by letting the model learn cross-stream information efficiently by focusing on when phase should influence magnitude recovery and when magnitude should regularize phase. CSF blocks allow magnitude and phase to exchange information for joint efficient modeling without collapsing into a single representation. Each CSF block uses a butterfly network \cite{Luo2020LatticeNet} that refines the streams and injects cross-stream information through four learned coefficients $(a_1,a_2,b_1,b_2)$. We use two CSF blocks.

\textbf{iii) Residual Connection and Synthesis:} After the last CSF block, the magnitude head predicts a residual correction $\Delta\mathbf{M}$. This residual form enables directly predicting the missing upper-band contents rather than requiring the network to reconstruct the whole spectrum from scratch.

Since raw phase values are noisy and difficult to predict directly \cite{yin2020phasen}, the phase branch predicts pseudo-real $(PR)$ and pseudo-imaginary $(PI)$ components through Layer Normalization (LN) and a Feed-Forward Network (FFN). The wrapped wideband phase is then recovered as $\Phi_{\mathrm{wb}}=\operatorname{atan2}(PI,PR)$ \cite{lu2024towards}. The reconstructed magnitude and phase are combined into a complex spectrogram and converted back to waveform by iSTFT.

\subsection{Design of Our HF-Focused Discriminators}
\label{subsec:dsdisc}

\textbf{Motivation:} The errors related to reconstructing higher frequencies (HF) are localized rather than temporally average case. Therefore, the most audible failures often occur as short upper-band events, including chirps, transient bursts, over-bright vowels, and smeared fricatives. \textit{Traditional discriminators \cite{mandel2023aero, abreu2024aeromamba, hauret2023eben, lu2024towards, liu2022neural} can dilute these short events through average pooling, HF hallucinations, and unrealistic objective scores, which leaves weak gradients for localized artifacts.}

To solve above issues, we design
two lightweight HF discriminators: \textbf{HFSD} controls the frame-wise HF-energy ratio through a differentiable barrier; and \textbf{WOSD} scores flatness, centroid, and rolloff through a learned weighted objective.

\subsubsection{\textbf{HF Safeguard Discriminator (HFSD)}}
\label{subsec:disc_hfsd}

\textbf{Motivation:} HFSD targets frequently repeated HF over-boost rather than penalizing every HF event. This distinction is important because brief HF structures, such as fricatives, stop releases, and sharp transients, are natural components of speech. Therefore, HFSD  monitors whether the upper band occupies an excessive fraction of the local spectral energy and whether this behavior persists over time. This resulting safeguard punishes artificial brightening, hiss, and metallic artifacts while preserving legitimate short-duration HF detail. 

\textit{For the first time, we design HFSD with a differentiable safeguard relaxation using Chance Constrained problems and optimization proposed by} \cite{erdougan2006ambiguous}. A pseudo-code is shown in Alg. \ref{alg:HFSD} and explained below.

\textbf{Rows {1}--{3}}: The generator's output $y$ is processed by reflection padding for window length $w$ and waveform length $T$. The positive-frequency bins $F$ is calculated using FFT size $n_{fft}$. A complex spectrogram $X$ is generated using Hann-windowed STFT for hop size $h$ and window length $w$.

\textbf{Rows  4--7}: Power $P$ is calculated. Then, HFSD  defines the portion of the spectrum considered ``high-frequency" as $f_0 = \left\lfloor(1-\rho)F\right\rfloor$, where $\rho$ = 0.25 is the fraction of $F$ = 513 bins designated as high frequency. Then the high-frequency set, $\mathcal{F}_{\mathrm{HF}} =\{f_0,\ldots,F-1\}.$ For example, when $\rho$ = 0.25 and $F$ = 513, the HFSD  focuses specifically on the frequency bins 384-512, where reconstruction errors and hallucinations are most likely to occur. For each time frame $\tau$, HFSD computes the total energy contained in the upper band as $E_{\mathrm{hf}}(\tau)$. In contrast, the total spectral energy of the frame is $E_{\mathrm{tot}}(\tau)$. Then HFSD forms the normalized ratio $q(\tau)$.

Because $q(\tau)$ is normalized by the total frame energy, it measures relative upper-band dominance rather than absolute signal amplitude.  For example, suppose $E_{\mathrm{hf}}$ = 20 and $E_{\mathrm{tot}}$ = 100, then $q$ = 0.2. It means that although the signal is loud-out, only 20\% of the energy lies in the upper band. HFSD sees this as normal. However, for $E_{\mathrm{hf}}$ = 15 and $E_{\mathrm{tot}}$ = 20, then $q$ = 0.75. Now 75\% of all energy resides in the upper band. This indicates an unnaturally bright or noisy reconstruction. HFSD strongly penalizes this case. \textit{Thus, a loud but spectrally balanced frame is not necessarily penalized, whereas a quieter frame with an unnaturally large HF proportion is punished harshly.}

\begin{algorithm}[ht!]
\small
\caption{Pseudo code of HFSD}
\label{alg:HFSD}
\begin{algorithmic}[1]
\REQUIRE $y_{\mathrm{real}},y_{\mathrm{fake}}$; $(n_{\mathrm{fft}}$=1024,h=80,w=320); $\rho,\kappa,\tau_0,\varepsilon$; $L_b$
\ENSURE $[r],[f],\{F_r^{(\ell)}\}_{\ell=1}^{2},\{F_f^{(\ell)}\}_{\ell=1}^{2}$
\STATE \textbf{Function ForwardSingleLoop (y)}
    \STATE $x\gets\operatorname{ReflectPad}\!\left(y,(w-T)_+\right)$
    \STATE $F\gets\lfloor n_{\mathrm{fft}}/2\rfloor+1,\quad
           X\gets\operatorname{STFT}(x;n_{\mathrm{fft}},h,w,\mathrm{Hann})$
    \STATE $P(f,\tau)\gets|X(f,\tau)|^2, f_0\gets\lfloor(1-\rho)F\rfloor$
    \STATE $E_{\mathrm{tot}}(\tau)\gets\sum_{f=0}^{F-1}P(f,\tau),E_{\mathrm{hf}}(\tau)\gets\sum_{f=f_0}^{F-1}P(f,\tau)$
    \STATE $q(\tau)\gets E_{\mathrm{hf}}(\tau)/E_{\mathrm{tot}}(\tau)$
    \STATE $b(\tau)=\kappa^{-1}\operatorname{softplus}
           \!\left(\kappa[q(\tau)-\tau_0]\right),
           h^{(0)}\gets b[:,,:]$
    \FOR{$\ell=1$ to $L_b$}
        \STATE $h^{(\ell)}\gets
        \operatorname{LeakyReLU}_{0.2}\!\left(
        \operatorname{BN}^{(\ell)}
        \left(\operatorname{DSC}^{(\ell)}
        (h^{(\ell-1)})\right)\right)$
    \ENDFOR
    \STATE $h_{\mathrm{out}}\gets
    \operatorname{BN}^{(L_b+1)}
    \left(\operatorname{DSC}^{(L_b+1)}(h^{(L_b)})\right)$
    \STATE $g_{\mathrm{avg}}\gets\operatorname{AdaptiveAvgPool1D}(h_{\mathrm{out}},1)$
           \STATE $g_{\mathrm{max}}\gets\operatorname{AdaptiveMaxPool1D}(h_{\mathrm{out}},1)$
    \STATE $s\gets\operatorname{squeeze}(g_{\mathrm{avg}}+g_{\mathrm{max}},-1)$
    \STATE $\{F^{(1)},F^{(2)}\}\gets\{h^{(0)},h_{\mathrm{out}}\}$
    \RETURN $(s,\{F^{(\ell)}\}_{\ell=1}^{2})$
\STATE $(r,\{F_r^{(\ell)}\})\gets\textbf{ForwardSingleLoop}({y_{\mathrm{real}}}),$
\STATE $(f,\{F_f^{(\ell)}\})\gets\textbf{ForwardSingleLoop}({y_{\mathrm{fake}}})$
\RETURN $([r],[f],\{F_r^{(\ell)}\},\{F_f^{(\ell)}\})$
\end{algorithmic}
\end{algorithm}

We define an acceptable upper limit $\tau_0$ = 0.4. If $q(\tau)>\tau_0$, the frame is considered to violate the safeguard. The empirical violation frequency is
$
\widehat{p}_{\mathrm{viol}}
=
\frac{1}{L}
\sum_{\tau=1}^{L}
\mathbb{I}\!\left\{q(\tau)>\tau_0\right\}
$, where $L$=number of frames and $\mathbb{I}$=indicator function. Here, $\mathbb{I}$=1 for $q(\tau)>\tau_0$ and $\mathbb{I}$=0 for otherwise. The $\widehat{p}_{\mathrm{viol}}$ estimates the fraction of frames that contain excessively HF energy. 
Please note that indicator function $\mathbb{I}$ is discontinuous. Its derivative is 0 almost everywhere and undefined at the threshold. Therefore, gradient-based training cannot optimize through it. To overcome this problem, HFSD replaces the hard indicator with the smooth positive-part approximation $
b(\tau)$, which is essentially a softplus barrier function. For example, if $q(\tau)<<\tau_0$, then $q(\tau)-\tau_0<0$, and $b(\tau)\approx0$. No penalty is applied. But when $q(\tau)\approx0$, the barrier smoothly increases. And when $q(\tau)>>\tau_0$, then $b(\tau)\approx q(\tau) - \tau_0.$ This means that the penalty grows almost linearly with the magnitude violation and therefore, large exceedances receive stronger punishment.

The sharpness parameter $\kappa$ = 8 controls how closely the barrier resembles a hard threshold. The $\kappa>0$ controls the transition sharpness. As $\kappa$ increases,
$b(\tau)$ approaches $\max(q(\tau)-\tau_0,0)$, which is the positive-part function. HFSD therefore becomes increasingly similar to a hard safeguard.

\textbf{Rows 8--11}: We process the barrier sequence through 4 convolution blocks ($L_b$=4). Each convolution block has an 1-D depth-wise separable convolution (DSC), batch normalization (BN), and LeakyReLU activation (see Fig. \ref{fig:overall_architecture}.c). 

\textbf{Rows 12--14}: We combine adaptive average and maximum pooling into $S$. Average pooling captures persistent HF abnormality, while maximum pooling retains sensitivity to the strongest localized artifact. Their sum therefore balances violation frequency and severity (see Fig. \ref{fig:overall_architecture}.c).

\textbf{Rows  15--19}: We  return
$
\{h^{(0)},h_{\mathrm{out}}\}
\label{eq:hfsd_feature_maps}
$
for feature matching. Matching $h^{(0)}$ aligns the HF-exceedance profiles of real and generated speech, whereas matching $h_{\mathrm{out}}$ aligns learned temporal patterns, such as violation duration and repetition. Then, we return the outputs for both real and generated signals to the interface required by the adversarial training.

\textbf{Generator-side losses:} Beyond adversarial discriminator, HFSD also regulates the amount of HF energy generated by the generator. For this purpose, the mean barrier $\bar b$ is calculated over frames and batches of $b$ and used as a generator constraint. Let $\bar b_{\mathrm{fake}}$ be evaluated on generated audio, and let $\bar b_0$ be its admissible target. The constraint residual is
$
g_{\mathrm{HF}}
=
\bar b_{\mathrm{fake}}-\bar b_0,
$ and measures the degree to which the generator exceeds the desired HF safeguard threshold. For example, if $g_{\mathrm{HF}}< 0$, the generator satisfies the safeguard; if $g_{\mathrm{HF}} = 0$, the generator operates exactly at the target level; and if $g_{\mathrm{HF}} > 0$, excessive HF dominance is present.

To enforce the safeguard during optimization, the residual $g_{\mathrm{HF}}$ is incorporated into an augmented-Lagrangian loss as
$
\mathcal{L}_{\mathrm{CC}}
=
\lambda_{\mathrm{cc}}g_{\mathrm{HF}}
+
\frac{\rho_{\mathrm{cc}}}{2}g_{\mathrm{HF}}^2.
\label{eq:hfsd_augmented_lagrangian}
$
The 1st term $\lambda_{\mathrm{cc}}g_{\mathrm{HF}}$ is a classical Lagrange multiplier, where $\lambda_{\mathrm{cc}}$ acts as a dual variable that dynamically adjusts the importance of the safeguard during training. The $\lambda_{\mathrm{cc}}$  is updated as
$
\lambda_{\mathrm{cc}}
\leftarrow
\operatorname{clip}\!\left(
\lambda_{\mathrm{cc}}
+
\eta_{\mathrm{cc}}g_{\mathrm{HF}},
0,
\lambda_{\max}
\right)
\label{eq:hfsd_dual_update}
$ with 0.01 step within 0-10. The 2nd term $\frac{\rho_{\mathrm{cc}}}{2}g_{\mathrm{HF}}^2$ introduces an extra penalty ($\rho_{\mathrm{cc}}$=0.5) that grows quadratically with the magnitude of the violation. During early training, $\mathcal{L}_{\mathrm{CC}}$ is scaled by the adversarial warm-up factor so that the safeguard does not dominate before the generator learns a plausible coarse reconstruction. 
Repeated positive residuals increase the safeguard pressure, whereas satisfied constraints prevent unnecessary growth. \textit{Therefore, HFSD does not suppress HF contents globally; it penalizes unnatural upper-band dominance while retaining valid fricatives, transients, and reconstructed harmonics successfully.}

\subsubsection{Weighted Objective Scoring Discriminator (WOSD)}
\label{subsec:disc_wosd}

\textbf{Motivation:}
We design WOSD to treat HF quality as an interpretable multi-criteria decision \cite{keeney1993decisions} rather than an energy-only score. It evaluates spectral flatness, normalized centroid, and rolloff so that noisiness, brightness, and spread are separated before aggregation. A pseudo-code is shown in Alg. \ref{alg:WOSD} and explained below.

\textbf{Rows  1--4}: The output $y$ from generator is similarly processed by reflection padding, and positive-frequency bins $F$ and a complex spectrogram $X$ is generated. The magnitude spectrum $A$ and total spectral energy $E(\tau)$ is calculated.

\textbf{Rows  5--8}: We compute three criteria -- spectral flatness ($\operatorname{SF}(\tau)$), normalized spectral centroid ($\operatorname{SC}(\tau)$), and $85\%$ spectral rolloff ($\operatorname{SR}_{85}(\tau)$). $\operatorname{SF}$ measures the geometric-to-arithmetic mean of magnitude spectrum. Intuitively, $\operatorname{SF}(\tau)$ quantifies whether a frame ($\tau)$ resembles a \textbf{(i)} tonal signal, where energy is concentrated in a few frequency bins, or a \textbf{(ii)} noise-like signal, where energy is distributed more uniformly. A low SF value indicates harmonic or tonal content, whereas a high SF indicates broadband spectral energy. \textbf{Since BWE often generates artificial noise-like artifacts, SF serves as an effective indicator of unnatural spectral distributions.}

$\operatorname{SC}(\tau)$ estimates the ``center of mass" of the spectrum. Higher SC indicates that energy is concentrated at higher frequencies, while lower values correspond to energy concentrated in lower-frequency regions. {Therefore, SC provides a direct measure of spectral brightness and HF emphasis.}

$\operatorname{SR}_{85}(\tau)$ characterizes how rapidly spectral energy accumulates across frequency. First, the cumulative spectral energy $C_{\mathrm{c}}(f,\tau)$ is computed. Then the algorithm identifies the frequency bin $b_p(\tau)$ at which 85\% of the total spectral energy has been accumulated, with p=0.85. The bin $b_p(\tau)$is normalized by F-1. Therefore, $\operatorname{SR}_{85}(\tau)$ captures the upper boundary of the dominant spectral energy distribution. \textbf{Signals containing excessive artificial high-frequency components generally exhibit larger $\operatorname{SR}_{85}(\tau)$ values because more energy resides in the upper spectral region.}

\begin{algorithm}[t!]
\caption{Pseudo code of WOSD}
\label{alg:WOSD}
\small
\begin{algorithmic}[1]
\REQUIRE $y_{\mathrm{real}},y_{\mathrm{fake}}$;
$(n_{\mathrm{fft}}$=1024,h=80,w=320); $f_s$; $p=0.85$; $L_u$
\ENSURE $[r],[f],\{F_r^{(\ell)}\}_{\ell=1}^{2},
\{F_f^{(\ell)}\}_{\ell=1}^{2}$

\STATE \textbf{Function ForwardSingleLoop (y)}
    \STATE $x\gets
    \operatorname{ReflectPad}\!\left(y,(w-T)_+\right)$

    \STATE $
    F\gets\lfloor n_{\mathrm{fft}}/2\rfloor+1,
    X\gets\operatorname{STFT}
    (x;n_{\mathrm{fft}},h,w,\mathrm{Hann})
    $

    \STATE $
    A\gets|X|+\varepsilon,
    E(\tau)\gets\sum_{f=0}^{F-1}A(f,\tau)+\varepsilon
    $

    \STATE $
        \operatorname{SF}(\tau)\gets
    \frac{
    \exp\!\left(F^{-1}\sum_f\log A(f,\tau)\right)}
    {F^{-1}\sum_fA(f,\tau)+\varepsilon}
    $

    \STATE $
        \operatorname{SC}(\tau)\gets
    \frac{\sum_f\nu_fA(f,\tau)}
    {E(\tau)(f_s/2+\varepsilon)}, \text{where}\, \nu_f\gets\operatorname{linspace}(0,f_s/2,F)
    $

    \STATE {$
        C_{\mathrm{c}}(f,\tau)\gets\sum_{j=0}^{f}A(j,\tau)
                $}

    \STATE $
        \operatorname{SR}_{85}(\tau)\gets
    \frac{b_p(\tau) = \sum_f\mathbf{1}
    \!\left[C_{\mathrm{c}}(f,\tau)<pE(\tau)\right] }{\max(1,F-1)}
    $

    \STATE $
         C(\tau)\gets
    [\operatorname{SF},\operatorname{SC},
    \operatorname{SR}_{85}]^\top, 
    C_n\gets\operatorname{IN}_{\mathrm{time}}(C)
    =
    \frac{C-\mu}{\sqrt{\sigma^2+\varepsilon}}
    $

    \STATE $
        \boldsymbol{\omega}\gets
    \operatorname{softmax}(\boldsymbol{\theta});
    u(\tau)\gets
    \sum_{k=1}^{3}\omega_kC_n(k,\tau);
    h^{(0)}\gets u[:,\mathrm{None},:]
    $

    \FOR{$\ell=1$ to $L_u$}

        \STATE $
                h^{(\ell)}\gets
        \operatorname{LeakyReLU}_{0.2}\!\left(
        \operatorname{BN}^{(\ell)}
        \!\left(
        \operatorname{DSC}^{(\ell)}
        (h^{(\ell-1)})
        \right)\right)
        $
    \ENDFOR

    \STATE $
        g\gets
    \operatorname{Flatten}\!\left(
    \operatorname{AdapAvgPool}
    (h^{(L_u)},1)\right);
    s\gets\operatorname{FC}(g)
    $

    \STATE $\{F^{(1)},F^{(2)}\}\gets\{C_n,u\}$

     \RETURN $(s,\{F^{(\ell)}\}_{\ell=1}^{2})$

\STATE $
(r,\{F_r^{(\ell)}\})\gets
\textbf{ForwardSingleLoop}({y_{\mathrm{real}}})
$

\STATE $
(f,\{F_f^{(\ell)}\})\gets
\textbf{ForwardSingleLoop}({y_{\mathrm{fake}}})
$
\RETURN $([r],[f],\{F_r^{(\ell)}\},\{F_f^{(\ell)}\})$
\end{algorithmic}
\end{algorithm}

\textbf{Rows 9--10}: The three spectral descriptors are fused as
$
C(\tau)$.
Next, $C(\tau)$ is normalized using mean ($\mu$) and variance ($\sigma^2$) along the temporal axis through instance normalization ($IN_{time}$). Then, WOSD introduces a learnable mechanism that automatically determines the importance of each spectral descriptor. A trainable parameter vector $\theta$ is converted into normalized weights $\omega$ through a softmax operation. The fused descriptor sequence is then computed as $\mu(\tau)$. This operation allows the network to adaptively emphasize the most discriminative spectral features during training. For example, if spectral centroid proves more informative than spectral flatness, the optimization process will naturally assign a larger weight to $SC(\tau$). The resulting 1-D sequence $\mu(\tau)$ is reshaped into a one-channel feature map $h^{(0)}$. 

\begin{table*}[ht]
\centering
\fontsize{9pt}{11pt}\selectfont
\setlength{\tabcolsep}{1.0pt}

\begin{tabular}{l |l |l|
  ccc | ccc | ccc | ccc | ccc | ccc}
\toprule
\multirow{2}{*}{Method}
& \multirow{2}{*}{Size}
& \multirow{2}{*}{Data}
  & \multicolumn{3}{c|}{NISQA-MOS$\uparrow$}
  & \multicolumn{3}{c|}{STOI$\uparrow$}
  & \multicolumn{3}{c|}{PESQ$\uparrow$}
    & \multicolumn{3}{c|}{SI-SNR$\uparrow$}
  & \multicolumn{3}{c|}{LSD$\downarrow$} 
  & \multicolumn{3}{c}{WER \%$\downarrow$}\\
  \cmidrule(lr){4-6}\cmidrule(lr){7-9}\cmidrule(lr){10-12}
\cmidrule(lr){13-15}\cmidrule(lr){16-18}\cmidrule(lr){19-21}
  & &
  & 4–16 & 8–16 & 16–48
  & 4–16 & 8–16 & 16–48
  & 4–16 & 8–16 & 16–48
    & 4–16 & 8–16 & 16–48
  & 4–16 & 8–16 & 16–48 
  & 4–16 & 8–16 & 16–48\\
\midrule
\multirow{2}{*} {Unprocessed}   & \multirow{2}{*} {-}     & VCTK & 2.79 & 3.67 & 4.43 & 0.55 & 0.61 & 0.61 & 1.15  & 1.51  & 1.41   & -10.53 & -7.62 & -5.63 & 3.27 & 2.27 & 2.85  & 90.0 & 6.1 & 2.1\\

&   & MLS  & 2.29 & 3.12 & - & 0.51 & 0.57 & - & 1.05  & 1.34  & -   & -13.67 & -17.71 & - & 3.48 & 2.48 & - & 91.1 & 6.8 & -\\

\hline

\multirow{2}{*} {EBEN}   & \multirow{2}{*} {29.7M} & VCTK & 2.59 & 2.69 & 2.53     & 0.89 & 0.98 & 0.98      & {\color{blue}2.64}  & 3.69  & 3.71            & 11.94  & 19.94 & 20.83      & 1.03  & 0.78  & 0.92   & 19.1 & 12.4 & 8.5 \\
   &   & MLS & 2.14 & 2.28 & -     & 0.87 & 0.96 & -      & 2.47  & 3.55  & -            & 11.74  & 18.38 & -      &  1.17 &  0.88 & -   & 19.4 & 12.7 & - \\

\hline

\multirow{2}{*} {AERO}  & \multirow{2}{*} {36.4M} & VCTK & 2.79 & 2.75 & 2.88 & 0.83 & 0.94 & 0.99  & 2.62  & 3.65  & 3.69   & {\color{blue}13.60}  & {\color{blue}20.70} & {\color{blue}21.56} & 1.09  & 0.97  & 0.75 & 20.3  & 13.4 & 9.8\\
&   & MLS  & 2.38 & 2.37 & - & 0.81 & 0.92 & -  & 2.49  & 3.45  & -   & {\color{blue}13.37}  & {\color{blue}19.19} & - & 1.21  & 1.04  & - & 20.7  & 13.7 & -\\

\hline

\multirow{2}{*} {AP-BWE} & \multirow{2}{*} {72M}  & VCTK & 3.86 & 3.97 & 4.49 & 0.94 & 0.99 & 0.99  & 2.55  & 3.69  & 3.72   & 13.35  & 18.07 & 20.74 & 0.96  & 0.74  & 0.75 & 13.7 & 4.2  & 1.7\\ 
&   & MLS & 3.18& 3.42 & - & 0.92&0.99  & -  & 2.51  & 3.62  & -  & 13.17  & 16.33 & - & 1.11  &  0.84 & - & 13.9 & 4.3 & -\\

\hline

\multirow{2}{*} {HFSWO}         & \multirow{2}{*} {18.3M} & VCTK & \textbf{4.32} & \textbf{4.40} & \textbf{4.52} & \textbf{0.94} & \textbf{0.99} & \textbf{0.99}  & \textbf{2.56}  & \textbf{3.70}  & \textbf{4.47}  & \textbf{13.37}  & \textbf{17.99} & \textbf{20.47} & \textbf{0.95}  & \textbf{0.73}  & \textbf{0.74} & \textbf{13.6} & \textbf{4.1} & \textbf{1.57}\\

(our) &   & MLS & \textbf{3.72} & \textbf{3.63} & - & \textbf{0.92} & \textbf{0.99} & -  & \textbf{2.52}  & \textbf{3.64} & -  & \textbf{13.16}  & \textbf{16.97} & - & \textbf{1.10}  & \textbf{0.78}  & - & \textbf{13.48} & \textbf{4.2} & \textbf{-}\\

\bottomrule
\end{tabular}
\caption{Clean-condition comparison on English VCTK and French MLS datasets; MLS does not include 48~kHz speech.}
\label{tab:Comparative_analysis}
\end{table*}

\textbf{Rows 11--14}:  $h$ is given as input to the four convolution blocks ($L_u$=4). Each block has a lightweight 1-D DSC, BN, and LeakyReLU activation (see Fig. \ref{fig:overall_architecture}.d).  Adaptive average pooling then summarizes the temporal evidence, and the fully connected (FC) layer produces the output $S$.

\textbf{Rows 15--19}: We return
$
\{C_n,u\}
\label{eq:wosd_feature_maps}
$
for feature matching. Matching $C_n$ aligns the individual flatness, centroid, and rolloff trajectories of real and generated speech, whereas matching $u$ aligns their learned convex aggregation. Then, we return the shared-path outputs, evaluate the real and generated signals separately, and package their scores and feature maps in the interface required by the adversarial training.

\textbf{Generator-side losses:} In addition to providing an adversarial real/fake score, WOSD returns intermediate feature representations that are used to construct a feature-matching (FM) loss. The feature-matching objective is defined as  $
\mathcal{L}_{\mathrm{FM}}^{\mathrm{WOSD}}$
=
$\sum_{\ell=1}^{2}
\frac{1}{N_\ell}
\left\|
F_r^{(\ell)}-F_f^{(\ell)}
\right\|_1,
\label{eq:wosd_feature_matching}
$
where $N_\ell$ is number of elements in $\ell$-th feature map,  and $F_r^{(\ell)}$ and $F_f^{(\ell)}$ correspond to the features extracted from real and generated waveforms, respectively. The FM loss is calculated in two feature levels. The 1st feature level corresponds to the normalized $C_{n}$ as $F^{(1)}$=$C_{n}$ and the 2nd feature level corresponds to the learned sequence $u$ as $F^{(2)}$=$u$ (see Row 15 of Alg. \ref{alg:WOSD}).

For the 1st feature level, the FM loss becomes:
$
\left\|C_{n,\mathrm{real}}-C_{n,\mathrm{fake}}\right\|_1.
$ This forces the generated signal to match the temporal change of the three spectral descriptors -- SF, SC, and $SR_{85}$.
For the 2nd feature level, the FM loss becomes:
$
\left\|u_{\mathrm{real}}-u_{\mathrm{fake}}\right\|_1.
$
This encourages the generator to reproduce not only individual descriptor values but also the relationships among them that WOSD has identified as important. \textit{Therefore, WOSD  provides interpretable HF feedback while preserving valid fricatives and transient spectral structure.} \textbf{{\color{black}See Technical Supplement
 8.7 and 9 for details.}}

\subsection{\textbf{Loss Functions and Training Objectives}
\label{subsec:losses}}

Lets assume MSE = mean squared loss, $C$ and $\hat{C}$ are the target and generated complex spectrograms, and $M$ and $\hat{M}$ are the target and generated magnitude from spectrograms. {\color{black}The complete generator objective is}
$
\mathcal{L}_{G}=45
\mathcal{L}_{\mathrm{mag}}
+100\mathcal{L}_{\mathrm{pha}}
+90\mathcal{L}_{\mathrm{com}}
+\gamma_{\mathrm{adv}}
\Big(
\mathcal{L}_{\mathrm{FM}}
+\mathcal{L}_{\mathrm{adv}}
\Big) 
+\mathcal{L}_{\mathrm{CC}},
$ and explained below.

The magnitude reconstruction loss, $\mathcal{L}_{\mathrm{mag}}$ = $\mathrm{MSE}(\hat{M},M)$. The phase reconstruction loss is
$\mathcal{L}_{\mathrm{pha}}
=
\left(
\mathcal{L}_{\mathrm{IP}}
+
\mathcal{L}_{\mathrm{GD}}
+
\mathcal{L}_{\mathrm{IAF}}
\right)
$,
where $\mathcal{L}_{\mathrm{IP}}$, $\mathcal{L}_{\mathrm{GD}}$, and
$\mathcal{L}_{\mathrm{IAF}}$ denote the instantaneous-phase, group-delay,
and instantaneous-amplitude-frequency losses, respectively, adopted from AP-BWE \cite{lu2024towards}. The complex STFT reconstruction loss is $\mathcal{L}_{\mathrm{com}}
=
\mathrm{MSE}(\hat{C},C)
$. 

Multi-Resolution Amplitude and Phase Discriminators (MRAD  and MRPD)  are adopted from \cite{lu2024towards} for lower bands with our newly designed high-frequency HFSD and WOSD discriminators.  The feature matching loss, $ \mathcal{L}_{\mathrm{FM}}$ =
$\sum_{d\in\mathcal{D}_{\mathrm{FM}}}\lambda_d
\sum_{\ell}\mathbb{E}\!\left[\left\lVert f^{\mathrm{real}}_{d,\ell}-f^{\mathrm{fake}}_{d,\ell}\right\rVert_1\right]$, where $\mathcal{D}_{\mathrm{FM}}$ = $\{\mathrm{MRAD},\mathrm{MRPD},\mathrm{HFSD},\mathrm{WOSD}\}$, $f_{d,\ell}$ is the layer $\ell$-th feature map from $d$, and $\lambda_d$ is the weight. The generator hinge loss, $\mathcal{L}_{\mathrm{adv}}  $=$\sum_{d\in\mathcal{D}_{\mathrm{hinge}}}\lambda_d\,
\mathbb{E}_{\hat{y}\sim p_G}\!\left[\max\!\big(0,1-D_d(\hat{y})\big)\right]$, where $\mathcal{D}_{\mathrm{hinge}}=\{\mathrm{MRAD},\mathrm{MRPD},\mathrm{HFSD},\mathrm{WOSD}\}$, $D_d(\cdot)$ is the scalar hinge critic score of discriminator $d$. The $\mathcal{L}_{\mathrm{CC}}$ is the HFSD augmented-Lagrangian penalty loss (see Section \ref{subsec:disc_hfsd}). Here, $\gamma_{\mathrm{adv}}\in[0,1]$ is the adversarial warmup factor. The total discriminator hinge loss is $\mathcal{L}_{D}$=$\sum_{d\in\mathcal{D}_{\mathrm{hinge}}}\mathcal{L}_{D}^{(d)}$. All $\lambda$'s are hyperparameters. \textbf{{\color{black}See Technical Supplement 8.8 and 8.9 for loss definitions and training procedures.}}

\subsection{\textbf{Dataset and Preprocessing}}
\label{subsec:Dataset_Preprocessing}

We use the English VCTK v0.92 \cite{yamagishi2019cstr} and MLS French \cite{Pratap2020MLSAL} datasets with cross-speaker evaluation for multispeaker and multilingual validation. VCTK is the primary benchmark with 114 speakers and 88{,}329 utterances. For MLS French, a 35-hour subset is used. Audio is converted to mono, silence-trimmed when metadata are available, and resampled to the target rate. {\color{black}\textbf{See Technical Supplement 5 and 8.1 for details}}. 

\subsection{\textbf{Hyperparameters and Configuration}}
\label{subsec:Hyperparameter_Configuration}

We train for 50 epochs using AdamW. The generator is initialized with a learning rate of $2\times10^{-4}$, and the discriminator learning rate is set to $3.5\times10^{-4}$. Both optimizers use $\beta_1=0.8$, $\beta_2=0.99$, and a weight-decay coefficient of 0.01. An exponential learning-rate scheduler with a decay factor of 0.999/epoch is applied. The effective batch size is 16. A random seed=1234 is used. Training is performed using an NVIDIA RTX 4090 GPU and an AMD Ryzen 9 7950X3D CPU. {\color{black}\textbf{See Technical Supplement 7, 8.2--8.6 for details.}}

\subsection{\textbf{Evaluation Metrics} }
\label{subsec:EvaluationMetrics}

We use complementary combination of objective and perceptual metrics. Log-spectral distance (LSD) \cite{115972} measures spectral deviation, while the short-time objective intelligibility  (STOI) \cite{taal2011algorithm} assesses speech intelligibility. Perceptual evaluation of speech quality (PESQ) \cite{rix2001perceptual} estimates perceived speech quality, and scale-invariant signal-to-noise ratio (SI-SNR) \cite{luo2018tasnet} quantifies waveform reconstruction quality independently of signal scale. We additionally report NISQA-MOS \cite{mittag2021nisqa} as a non-intrusive estimate of perceptual quality. To determine whether bandwidth extension also benefits downstream recognition, we also evaluate word error rate (WER). \textit{To calculate PESQ and STOI over 16 kHz, we first downsample the signals to 16 kHz.}

\section{Results and Analysis}
\label{sec:Results and Analysis}

\subsection{Comparative Analysis with Baselines}
\label{subsec:ComparativeAnalysiswithBaselines}

Table~\ref{tab:Comparative_analysis} compares HFSWO with three most recent baselines: EBEN~\cite{hauret2023eben}, which employs a pseudo-QMF architecture; AERO~\cite{mandel2023aero}, which performs complex-valued spectral reconstruction; and AP-BWE~\cite{lu2024towards}, which jointly predicts amplitude and phase through dual processing streams. The comparison covers the 4$\rightarrow$16, 8$\rightarrow$16, and 16$\rightarrow$48~kHz extension settings on the English VCTK and MLS French datasets. 

On both VCTK and MLS datasets, HFSWO provides significantly improved NISQA-MOS over all three recent baselines. Moreover, HFSWO provides improved STOI, PESQ, LSD, and WER over all the recent baselines for all three extension ranges (except PESQ for EBEN at 4-16 kHz). However, HFSWO provides slightly lower SI-SNR compared to AERO, mixed SI-SNR results compared to AP-BWE and EBEN. Although its SI-SNR values do not surpass every competing system, the combined NISQA-MOS, PESQ, STOI, and WER show that its reconstruction is perceptually effective and preserves linguistic content at the widest extension range in multilingual and multispeaker settings.

\vspace{-0.00em}
\textbf{Superior Quality with a Fraction of Parameter:} HFSWO consistently provides better performance with only 18.5 M parameters, which is approximately 74.6\% less than the most competing baseline AP-BWE (72 M). The results indicate that HFSWO dominates most metrics with substantially lower parameter than all recent baselines.

\begin{figure}[ht!]
  \centering
 \includegraphics[width=0.48\textwidth,height=0.17
 \textheight]{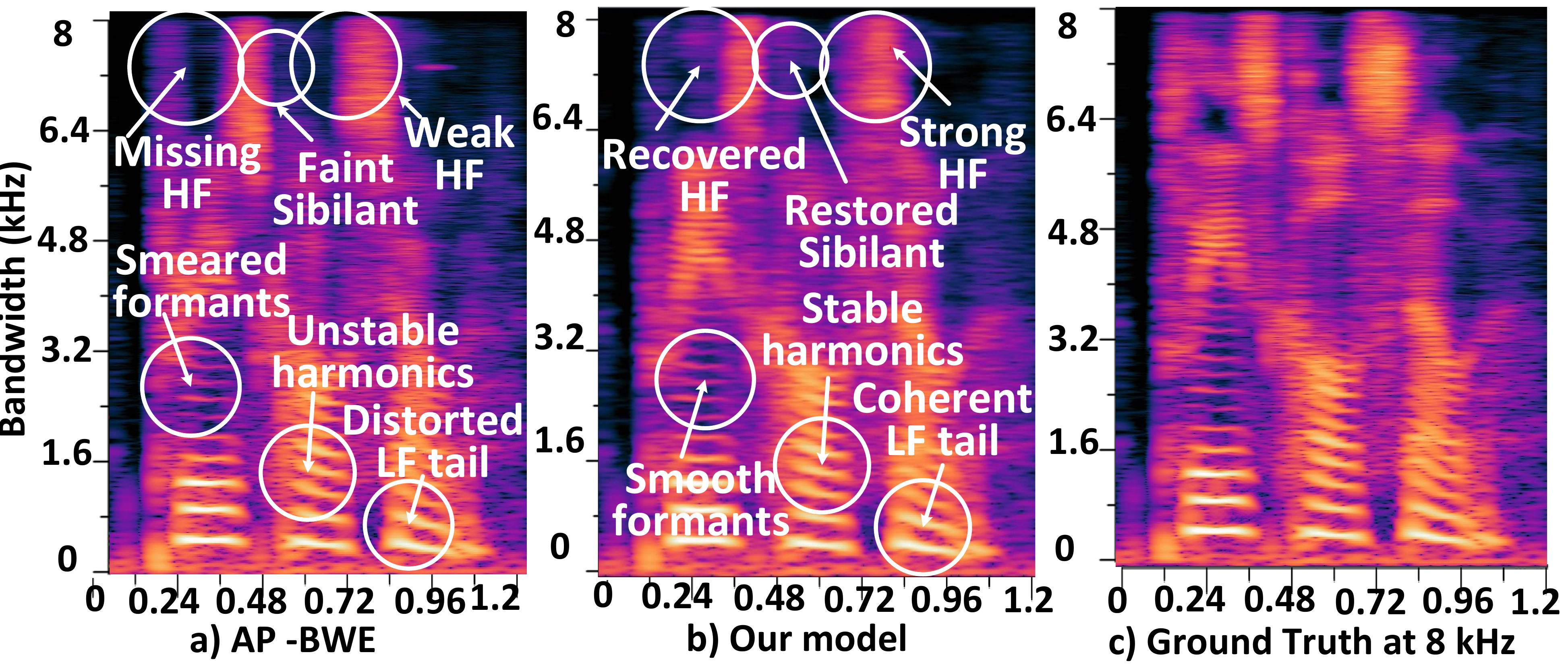} 

  \caption{{\color{black}Proof of recovered HF components for 4--16~kHz.}}
  \label{fig:figcomparison}
  \vspace{-0em}
\end{figure}

\textbf{Qualitative analysis:} As shown in Fig.~\ref{fig:figcomparison}(a), AP-BWE only partially reconstructs the missing upper-frequency region: several high-frequency components remain weak or absent, and the sibilant structure around 6.4--8~kHz is faint. The reconstruction also exhibits smeared formant trajectories, unstable harmonic patterns in the mid-frequency region, and a distorted low-frequency (LF) tail. In contrast, Fig.~\ref{fig:figcomparison}(b) shows that HFSWO recovers a denser and more continuous upper-band representation. The reconstructed sibilant energy is more clearly defined, and the high-frequency components more closely follow the temporal locations and spectral shapes observed in the ground truth in Fig.~\ref{fig:figcomparison}(c).

\subsection {\textbf{ITU-Derived Evidence in HF Reconstruction}}
\label{subsec:itu_inspired_hfswo_evidence}

For the first time, we provide results to support our HFSWO's HF reconstruction accuracy using a sub-band analysis by three ITU (International Telecommunication Union)-derived diagnostic metrics: \textbf{1.} the band-energy-ratio error $\mathrm{ERE}_{i}$ derived from ITU-R BS.1387 \cite{bs1387_2023}; \textbf{2.} the K-weighted loudness error $D_{\mathrm{KLU},i}$ derived from ITU-BS1770 \cite{itu_bs1770_2023}; and \textbf{3.} the active-speech energy-ratio error $D_{\mathrm{P56},i}$ derived from P.56 \cite{itu_p56_2011}. These metrics provide complementary properties -- spectral energy allocation, perceived loudness consistency, and active-speech energy distribution of reconstructed speech. All three metrics are error measures; hence, lower values indicate closer agreement with the ground-truth audio. {\color{black}\textbf{We refer to Technical Supplement 6.2 for details on these metrics.}}

\begin{table}[ht!]
\centering
\fontsize{9pt}{11pt}\selectfont
\setlength{\tabcolsep}{1.4pt}
\begin{tabular}{lcccccc}
\toprule

Band &
\multicolumn{2}{c}{$\mathrm{ERE}_{i}$ $\downarrow$} &
\multicolumn{2}{c}{$D_{\mathrm{KLU},i}$ $\downarrow$} &
\multicolumn{2}{c}{$D_{\mathrm{P56},i}$ $\downarrow$}  \\
\cmidrule(lr){2-3}\cmidrule(lr){4-5}\cmidrule(lr){6-7}
kHz & AP-BWE & HFSWO
& AP-BWE & HFSWO
& AP-BWE & HFSWO \\
\midrule

0--4 
& 0.002688 & \textbf{0.002536}
& 0.058 & \textbf{0.035}
& 0.002722 & \textbf{0.002557} \\

4--8 
& 0.001415 & \textbf{0.000623}
& 0.277 & \textbf{0.199}
& 0.001387 & \textbf{0.000578} \\

8--12 
& 0.002154 & \textbf{0.002077}
& \textbf{2.232} & 2.246
& 0.001994 & \textbf{0.001956} \\

\bottomrule
\end{tabular}
\caption{Sub-band analysis using ITU-derived evaluation metrics as evidence of HF reconstruction improvement.}

\label{tab:all_subband_selected_metrics}
\end{table}

Table~\ref{tab:all_subband_selected_metrics} shows that
HFSWO obtains the better (lower) error in most comparisons. The most pronounced evidence appears for the 4--8 kHz band, which lies immediately below the missing HF region. In this band, HFSWO improves $\mathrm{ERE}_{i}$ by 56\%, $D_{\mathrm{KLU},i}$ by 28.2\%, and $D_{\mathrm{P56},i}$ by 58.4\%. The overall pattern therefore supports that the proposed HFSWO is not simply increasing HF magnitude; it is reconstructing HF speech with more coherent temporal organization. This behavior is precisely what is needed for perceptually important cues such as fricative bursts, sibilant onsets, and rapidly varying HF speech envelopes.

\subsection{\textbf{Evaluation under Noisy Conditions}}
\label{subsec:noisy_conditions}

Noisy evaluation is performed using eight different noises from AURORA dataset \cite{pearce00_icslp} at five SNR levels ($-10$, $-5$, $0$, $5$, and $10$~dB), producing 40 different noise--SNR combinations. The noise set comprises airport, babble, car, exhibition, station, street, speech-shaped noise (SSN), and additive white Gaussian noise (AWGN), covering stationary and non-stationary interference. Table~\ref{tab:noisy_snr_wise_detail} compares the best-performing baseline AP-BWE  with our HFSWO. \textbf{HFSWO obtains higher NISQA-MOS and SI-SNR at every SNR level for all three extension settings.}

\begin{table}[t!]
\centering
\fontsize{9pt}{9pt}\selectfont
\setlength{\tabcolsep}{3.0pt}
\begin{tabular}{ ll  ccc ccc}
\toprule

\multirow{2}{*}{Model}
& \multirow{2}{*}{SNR}
& \multicolumn{3}{c}{NISQA-MOS$\uparrow$}
& \multicolumn{3}{c}{SI-SNR$\uparrow$}\\
\cmidrule(lr){3-5}
\cmidrule(lr){6-8}
 & &
  4--16 & 8--16 & 16--48
& 4--16 & 8--16 & 16--48 \\
\midrule

\multirow{5}{*}{AP-BWE}
& -10 
& 2.02 & 2.55 & 2.98
& -0.60 & -0.04 & 1.17 \\

 & -5 
& 2.52 & 2.86 & 3.54
& 2.86 & 3.48 & 4.44\\

 & 0 
& 2.83 & 3.21 & 3.80
& 5.06 & 5.83 & 6.80\\

 & 5 
& 3.05 & 3.42 & 4.07
& 6.53 & 7.42 & 8.64\\

 & 10 
& 3.22 & 3.70 & 4.22
& 7.62 & 9.02 & 9.77\\

\midrule

 \multirow{5}{*}{HFSWO}
& -10 
& 2.51 & 3.01 & 3.37
& -0.29 & 0.30 & 1.29\\

 & -5 
& 3.17 & 3.56 & 3.83
& 3.15 & 3.58 & 4.52\\

 & 0 
& 3.76 & 3.93 & 4.13
& 5.31 & 5.87 & 7.25\\

 & 5 
& 4.14 & 4.19 & 4.32
& 6.58 & 7.44 & 9.32\\

 & 10 
& 4.31 & 4.35 & 4.44
& 7.63 & 9.03 & 10.77\\

\bottomrule
\end{tabular}
\caption{Expanded SNR-wise results on noisy VCTK dataset.}
\label{tab:noisy_snr_wise_detail}
\end{table}

\subsection{\textbf{Computational Complexity}}
\label{subsec:ncomp_comp}

Table~\ref{tab:computation} reports the model complexity and inference efficiency using total parameters, \textit{MAC} (multiply--accumulate operations), \textit{FLOP} (floating-point operations), \textit{RTF} (real-time factor), and inference \textit{Latency}. HFSWO has 74.6\% fewer parameters (18.3 M vs 72.07 M) and 63.6\% less MAC and FLOP (5177 M vs 14236 M) compared to the best-performing AP-BWE. However, HFSWO exhibits similar/slightly higher latency (15.54--16.98~ms) and RTF (0.0027--0.0029$\times$) compared to AP-BWE. As HFSWO's  RTF is substantially less than 1, it is adequately suitable for real-time operations. 
\begin{table}[ht!]
\centering
\fontsize{9pt}{11pt}\selectfont
\setlength{\tabcolsep}{1.30pt}

\begin{tabular}{ c c c c c c c}
\toprule
Model & Freq. & Param. & MAC(M) & FLOP(M) & RTF(M) & Latency \\
\midrule
 AP-BWE
& 4--16
& 72.07 M
& 14236.65
& 28473.31
& 0.0023$\times$
& 14.93ms\\

 HFSWO
& 4--16
& \textbf{18.30 M}
& \textbf{5177.89}
& \textbf{10355.77}
& \textbf{0.0027$\times$}
& \textbf{15.54ms} \\
\hline\hline
 HFSD & 4--16 &7.6 k
& 0.141
& 0.282
& x
& 0.823 \\

 WOSD & 4--16 & 7.4 k
& 0.139
& 0.278
& x
& 0.889 \\
\bottomrule
\end{tabular}
\caption{Model- and discriminator-wise  complexity on GPU. Parameters, MAC, and FLOP are reported in millions (M).}
\label{tab:computation}
\end{table}

\textbf{Discriminator-only complexity:} Table~\ref{tab:computation} also reports the training-only complexity related to only the HFSD and WOSD discriminators. HFSD and WOSD each require approximately 0.14 M MACs, 0.28 M FLOPs, 7.6 k parameters, and less than 0.9~ms latency. These indicate that the training-time burden for HFSD and WOSD is trivial.

\subsection{\textbf{Subjective Mean Opinion Score (MOS)}}
\label{subsec:Subjective Test}

We conduct a listening test to evaluate HFSWO against AP-BWE. Ten English-speaking  volunteers (four male and six female) with self-reported normal hearing participate. Before scoring, listeners are instructed on the MOS scale,  and use soundproof headsets to standardize listening conditions. Each listener rates 30 sets of samples. Clips are 5--7~s long and are drawn from VCTK under 2--16~kHz, 12--48~kHz, and 24--48~kHz conditions. \textit{Our HFSWO, having MOS=4.32, outperforms AP-BWE, which has MOS=4.21.}

\subsection{\textbf{Statistical Analysis of MOS Scores}}
\label{subsec:nisqa_stats}

We compute statistical analysis on MOS scores over a 1,000-utterance matched subset of AP-BWE and HFSWO outputs using mean, standard deviation (SD), standard error of mean (SEM), 95\% confidence interval (CI), and paired Wilcoxon signed-rank test, tabulated in Tables \ref{tab:nisqa_mos_stats_1000} and  \ref{tab:wilcoxon_nisqa_1000}.

HFSWO shows higher mean MOS with a higher 95\% confidence interval over AP-BWE. The utterance-level Wilcoxon test indicates a statistically significant paired difference (\(p=5.04\times10^{-8}\)). In contrast, the speaker-level comparison does not reach statistical significance (p=0.469). This difference suggests that the utterance-level analysis is influenced by the multiple utterances contributed by each speaker, whereas the speaker-level analysis treats speakers as independent observations. Therefore, the results provide evidence of an utterance-level improvement but not sufficient evidence of a statistically significant improvement at the speaker level.

\begin{table}[ht!]
\centering
\fontsize{9pt}{11pt}\selectfont
\setlength{\tabcolsep}{3.7pt}
\begin{tabular}{llccccc}
\hline
Agg. & Method & $N$ & MOS $\uparrow$ & SD & SEM & 95\% CI \\
\hline
Utt. & AP-BWE    & 1000 & 4.138 & 0.593 & 0.019 & [4.101, 4.175] \\
Utt. & HFSWO & 1000 & \textbf{4.206} & 0.648 & 0.021 & \textbf{[4.166, 4.247]} \\
Spk. & AP-BWE    & 7    & 4.126 & 0.281 & 0.106 & [3.867, 4.386] \\
Spk. & HFSWO & 7    & \textbf{4.189} & 0.399 & 0.151 & \textbf{[3.821, 4.558]} \\
\hline
\end{tabular}
\caption{MOS descriptive statistics for the 1,000-utterance subset. Utt. = utterance-level; Spk. = speaker-level.}
\label{tab:nisqa_mos_stats_1000}

\end{table}

\begin{table}[ht!]
\centering
\fontsize{9pt}{11pt}\selectfont
\setlength{\tabcolsep}{0.0pt}
\setlength{\tabcolsep}{4pt}
\begin{tabular}{lccccc}
\hline
Agg.   & Mean $\Delta$ & Med. $\Delta$ & $W^+$ & $W^-$  & $p$ \\
\hline
Utt.   & 0.069 & 0.082 & 300{,}039 & 200{,}461  & $5.04\times10^{-8}$ \\
Spk.       & 0.063 & 0.094 & 19 & 9  & 0.469 \\
\hline
\end{tabular}
\caption{Paired Wilcoxon signed-rank test for NISQA-MOS differences. Mean $\Delta$/Med. $\Delta$ = mean/median paired difference; \(W^+\)/\(W^-\) = positive/negative rank sums.}
\label{tab:wilcoxon_nisqa_1000}
\end{table}

\subsection{\textbf{Ablation Study}}
\label{subsec:Ablation_Study}

Table \ref{table:Ablation_Study} evaluates the contribution of our proposed discriminators HFSD and WOSD along with existing MRAD and MRPD from AP-BWE \cite{lu2024towards}. Table \ref{table:Ablation_Study} indicates that including HFSD and WOSD on top of MRAD and MRPD improves all metrics, and that combining the four discriminators yields the best results. These gains support a two-level design: MRAD and MRPD maintain global spectral and temporal structure, whereas HFSD and WOSD additionally refine the reconstructed high-frequency region. Therefore, the discriminator suite is designed to balance complementary attributes rather than optimize a single metric.

\textbf{Generator–discriminator compatibility:} To study the individual performance of the generator, we connect all four discriminators to our Swin-based and AP-BWE's generator separately. Results are shown in the last two rows in Table \ref{table:Ablation_Study}, indicating that our Swin-based generator is more compatible.

\begin{table}[ht!]
\centering
\fontsize{9pt}{10pt}\selectfont
\setlength{\tabcolsep}{1.350pt}
\begin{tabular}{c c c c | c  | c | c | c}
\toprule
 MRAD & MRPD & HFSD & WOSD & LSD$\downarrow$  & PESQ$\uparrow$ & SI-SNR$\uparrow$ & NISQA$\uparrow$ \\
\midrule

 \cmark & \xmark & \cmark & \xmark & 0.97  & 2.35 & 12.9 & 4.28 \\
 \cmark & \cmark & \cmark & \xmark & 1.01  & 2.27 & 12.99 & 4.29\\

 \cmark & \cmark & \xmark & \cmark & 0.99  & 2.3 & 12.65 & 4.28 \\
 \cmark & \cmark & \cmark & \cmark & \textbf{0.95}  & \textbf{2.56} & \textbf{13.37} & \textbf{4.32}\\
 \midrule
\midrule

 \cmark & \cmark & \cmark & \cmark & 0.99& 2.29& 12.69& 4.26\\
 \cmark & \cmark & \cmark & \cmark & \textbf{0.95}  & \textbf{2.56} & \textbf{13.37} & \textbf{4.32}\\
\bottomrule
\end{tabular}
\caption{Ablation study on discriminators for 4$\rightarrow$16~kHz.} \label{table:Ablation_Study}
\end{table}

\section{Limitations and Conclusion}
\label{sec:limitation}

HFSWO has higher training time than AP-BWE (17 min. vs. 20 min./epoch) because of the added extra discriminators (see Table \ref{tab:computation}). Current design does not use data-driven calibration of hyperparameters in HFSD and WOSD.

We design the HFSWO BWE framework, for the first time, to explicitly control over where, how often, and in what spectral form HF energy should be produced to suppress unintended HF overboosting. Extensive experiments on VCTK English and MLS French datasets across multiple frequency ranges, including noisy conditions, show significantly improved perceptual scores with 74.6\% fewer parameters than the existing baseline AP-BWE, establishing a new baseline in the BWE task in multilingual and multispeaker setups

\bibliography{aaai2027}

@incollection{Luo2020LatticeNet,
  author       = {Xiaotong Luo and Yuan Xie and Yulun Zhang and Yanyun Qu and Cuihua Li and Yun Fu},
  title        = {{LatticeNet: Towards Lightweight Image Super‑Resolution with Lattice Block}},
  booktitle    = {Computer Vision – ECCV 2020 (Lecture Notes in Computer Science, vol 12367)},
  editor       = {Andrea Vedaldi and Horst Bischof and Thomas Brox and Jan‑Michael Frahm},
  pages        = {272--289},
  publisher    = {Springer, Cham},
  year         = {2020},
  doi          = {10.1007/978-3-030-58542-6_17},
}

@article{lu2024towards,
  title={Towards high-quality and efficient speech bandwidth extension with parallel amplitude and phase prediction},
  author={Lu, Ye-Xin and Ai, Yang and Du, Hui-Peng and Ling, Zhen-Hua},
  journal={IEEE/ACM Transactions on Audio, Speech, and Language Processing},
  year={2024},
  publisher={IEEE}
}

@inproceedings{pearce00_icslp,
  author    = {David Pearce and Hans-G{\"u}nter Hirsch},
  title     = {The Aurora Experimental Framework for the Performance Evaluation of Speech Recognition Systems under Noisy Conditions},
  booktitle = {Proceedings of the 6th International Conference on Spoken Language Processing (ICSLP 2000)},
  year      = {2000},
  pages     = {29--32},
  doi       = {10.21437/ICSLP.2000-743},
  publisher = {ISCA Archive},
}

@inproceedings{mandel2023aero,
  title={Aero: Audio super resolution in the spectral domain},
  author={Mandel, Moshe and Tal, Or and Adi, Yossi},
  booktitle={ICASSP 2023-2023 IEEE International Conference on Acoustics, Speech and Signal Processing (ICASSP)},
  pages={1--5},
  year={2023},
  organization={IEEE}
}

@article{wang2021towards,
  title={Towards robust speech super-resolution},
  author={Wang, Heming and Wang, DeLiang},
  journal={IEEE/ACM transactions on audio, speech, and language processing},
  volume={29},
  pages={2058--2066},
  year={2021},
  publisher={IEEE}
}

@inproceedings{lee2025wave,
  title={Wave-u-mamba: an end-to-end framework for high-quality and efficient speech super resolution},
  author={Lee, Yongjoon and Kim, Chanwoo},
  booktitle={ICASSP 2025-2025 IEEE International Conference on Acoustics, Speech and Signal Processing (ICASSP)},
  pages={1--5},
  year={2025},
  organization={IEEE}
}

@inproceedings{lee2021nu,
  title     = {NU-Wave: A Diffusion Probabilistic Model for Neural Audio Upsampling},
  author    = {Junhyeok Lee and Seungu Han},
  year      = {2021},
  booktitle = {Interspeech 2021},
  pages     = {1634--1638},
  doi       = {10.21437/Interspeech.2021-36},
  issn      = {2958-1796},
}

@inproceedings{yin2020phasen,
  title={Phasen: A phase-and-harmonics-aware speech enhancement network},
  author={Yin, Dacheng and Luo, Chong and Xiong, Zhiwei and Zeng, Wenjun},
  booktitle={Proceedings of the AAAI Conference on Artificial Intelligence},
  volume={34},
  pages={9458--9465},
  year={2020}
}

@article{abreu2024aeromamba,
  title={AEROMamba: An efficient architecture for audio super-resolution using generative adversarial networks and state space models},
  author={Abreu, Wallace and Biscainho, Luiz Wagner Pereira},
  journal={arXiv preprint arXiv:2411.07364},
  year={2024}
}

@inproceedings{rix2001perceptual,
  title={Perceptual evaluation of speech quality (PESQ)-a new method for speech quality assessment of telephone networks and codecs},
  author={Rix, Antony W and Beerends, John G and Hollier, Michael P and Hekstra, Andries P},
  booktitle={2001 IEEE international conference on acoustics, speech, and signal processing. Proceedings (Cat. No. 01CH37221)},
  volume={2},
  pages={749--752},
  year={2001},
  organization={IEEE}
}

@article{yamagishi2019cstr,
  title={CSTR VCTK Corpus: English multi-speaker corpus for CSTR voice cloning toolkit (version 0.92)},
  author={Yamagishi, Junichi and Veaux, Christophe and MacDonald, Kirsten and others},
  journal={University of Edinburgh. The Centre for Speech Technology Research (CSTR)},
  pages={271--350},
  year={2019}
}

@article{ashish2017attention,
  title={Attention is all you need},
  author={Ashish, Vaswani},
  journal={Advances in neural information processing systems},
  volume={30},
  pages={I},
  year={2017}
}

@inproceedings{kim2023hifi++,
  title={{HiFi++}: Towards Perceptually Enhanced and Computationally Efficient Neural Vocoders},
  author={Kim, Woosung and Kang, Hyeongmin and Kim, Younggun and Jung, Kyunggu and Lee, Joon Son and Lee, Sang-Hoon},
  booktitle={Proc. Interspeech},
  year={2023},
  pages={4374--4378}
}

@inproceedings{hauret2023eben,
  title={EBEN: Extreme Bandwidth Extension Network applied to speech signals captured with noise-resilient body-conduction microphones},
  author={Hauret, Julien and Joubaud, Thomas and Zimpfer, V{\'e}ronique and Bavu, {\'E}ric},
  booktitle={ICASSP 2023 - 2023 IEEE International Conference on Acoustics, Speech and Signal Processing (ICASSP)},
  pages={1--5},
  year={2023},
  organization={IEEE},
  doi={10.1109/ICASSP49357.2023.10096301}
}

@article{Pratap2020MLSAL,
  title        = {MLS: {A} Large-Scale Multilingual Dataset for Speech Research},
  author       = {Vineel Pratap and Qiantong Xu and Anuroop Sriram and Gabriel Synnaeve and Ronan Collobert},
  journal      = {arXiv preprint arXiv:2012.03411},
  year         = {2020},
  url          = {https://arxiv.org/abs/2012.03411},
}

@article{kumar2019melgan,
  title={Melgan: Generative adversarial networks for conditional waveform synthesis},
  author={Kumar, Kundan and Kumar, Rithesh and De Boissiere, Thibault and Gestin, Lucas and Teoh, Wei Zhen and Sotelo, Jose and De Brebisson, Alexandre and Bengio, Yoshua and Courville, Aaron C},
  journal={Advances in neural information processing systems},
  volume={32},
  year={2019}
}

@article{kong2020hifi,
  title={Hifi-gan: Generative adversarial networks for efficient and high fidelity speech synthesis},
  author={Kong, Jungil and Kim, Jaehyeon and Bae, Jaekyoung},
  journal={Advances in neural information processing systems},
  volume={33},
  pages={17022--17033},
  year={2020}
}

@article{taal2011algorithm,
  title={An algorithm for intelligibility prediction of time--frequency weighted noisy speech},
  author={Taal, Cees H and Hendriks, Richard C and Heusdens, Richard and Jensen, Jesper},
  journal={IEEE Transactions on audio, speech, and language processing},
  volume={19},
  number={7},
  pages={2125--2136},
  year={2011},
  publisher={IEEE}
}

@inproceedings{luo2018tasnet,
  title={Tasnet: time-domain audio separation network for real-time, single-channel speech separation},
  author={Luo, Yi and Mesgarani, Nima},
  booktitle={2018 IEEE International Conference on Acoustics, Speech and Signal Processing (ICASSP)},
  pages={696--700},
  year={2018},
  organization={IEEE}
}

@article{mittag2021nisqa,
  title={NISQA: A deep CNN-self-attention model for multidimensional speech quality prediction with crowdsourced datasets},
  author={Mittag, Gabriel and Naderi, Babak and Chehadi, Assmaa and M{\"o}ller, Sebastian},
  journal={arXiv preprint arXiv:2104.09494},
  year={2021}
}

@INPROCEEDINGS{115972,
  author={Erell, A. and Weintraub, M.},
  booktitle={International Conference on Acoustics, Speech, and Signal Processing}, 
  title={Estimation using log-spectral-distance criterion for noise-robust speech recognition}, 
  year={1990},
  volume={},
  number={},
  pages={853-856 vol.2},
  doi={10.1109/ICASSP.1990.115972}}

@article{titze2008nonlinear,
  title={Nonlinear source--filter coupling in phonation: Theory},
  author={Titze, Ingo R},
  journal={The Journal of the Acoustical Society of America},
  volume={123},
  number={5},
  pages={2733--2749},
  year={2008},
  publisher={AIP Publishing}
}

@inproceedings{liu2021swin,
  title={Swin transformer: Hierarchical vision transformer using shifted windows},
  author={Liu, Ze and Lin, Yutong and Cao, Yue and Hu, Han and Wei, Yixuan and Zhang, Zheng and Lin, Stephen and Guo, Baining},
  booktitle={Proceedings of the IEEE/CVF international conference on computer vision},
  pages={10012--10022},
  year={2021}
}

@inproceedings{liu2022neural,
  title={Neural vocoder is all you need for speech super-resolution},
  author={Liu, Haohe and Choi, Woosung and Liu, Xubo and Kong, Qiuqiang and Tian, Qiao and Wang, DeLiang},
  year      = {2022},
  booktitle = {Interspeech},
}

@book{keeney1993decisions,
  title={Decisions with multiple objectives: preferences and value trade-offs},
  author={Keeney, Ralph L and Raiffa, Howard},
  year={1993},
  publisher={Cambridge university press}
}

@article{tamiti2025nldsi,
  title={NLDSI-BWE: Non Linear Dynamical Systems-Inspired Multi Resolution Discriminators for Speech Bandwidth Extension},
  author={Tamiti, Tarikul Islam and Barua, Anomadarshi},
  journal={arXiv preprint arXiv:2510.01109},
  year={2025}
}

@article{erdougan2006ambiguous,
  title={Ambiguous chance constrained problems and robust optimization},
  author={Erdo{\u{g}}an, Emre and Iyengar, Garud},
  journal={Mathematical Programming},
  volume={107},
  number={1},
  pages={37--61},
  year={2006},
  publisher={Springer}
}

@techreport{bs1387_2023,
  title={Recommendation ITU-R BS.1387-2: Method for objective measurements of perceived audio quality},
  author={{ITUBS1387}},
  institution={{International Telecommunication Union}},
  type={Recommendation ITU-R},
  number={BS.1387-2},
  year={2023},
  month={November},
  address={Geneva, Switzerland},
  url={https://www.itu.int/rec/R-REC-BS.1387/en}
}

@techreport{itu_bs1770_2023,
  title={Recommendation ITU-R BS.1770-5: Algorithms to measure audio programme loudness and true-peak audio level},
  author={{ITUBS1770}},
  institution={{International Telecommunication Union}},
  type={Recommendation ITU-R},
  number={BS.1770-5},
  year={2023},
  month={November},
  address={Geneva, Switzerland},
  url={https://www.itu.int/rec/R-REC-BS.1770/en}
}

@techreport{itu_p56_2011,
  title={Recommendation ITU-T P.56: Objective measurement of active speech level},
  author={{ITUP56}},
  institution={{International Telecommunication Union}},
  type={Recommendation ITU-T},
  number={P.56},
  year={2011},
  month={December},
  address={Geneva, Switzerland},
  url={https://www.itu.int/rec/T-REC-P.56/en}
}

\end{document}